\documentstyle[preprint,axodraw,amsfonts,aps]{revtex}
\begin{document}
\title{Vacuum Polarization Effects in the Lorentz and PCT Violating Electrodynamics}
\author{G. Bonneau$^{[a]}$\footnote{{\it E-mail addresses} : bonneau@lpthe.jussieu.fr  (G. Bonneau);  
lccosta@ift.unesp.br (L.C. Costa)},  L. C. Costa$^{[a][b]}$ and J. L. Tomazelli$^{[a][c]}$ } 
\address{
$^{[a]}$ Laboratoire de Physique Th\'eorique et Hautes Energies\footnote{Unit\'e associ\'ee 
au CNRS UMR 7589.}, Universit\'es de Paris VI et Paris VII, \\  
2 Place Jussieu, 75251 Paris Cedex 05, France. \\
$^{[b]}$ On leave from Departamento de F\'{\i}sica, Universidade Federal de S\~ao Carlos, \\
P.O. Box 676, 13565-905, S\~ao Carlos, SP, Brazil. \\
$^{[c]}$ On leave from Departamento de F\'{\i}sica e Qu\'{\i}mica, Universidade Estadual Paulista, \\
Av. Dr. Ariberto Pereira da Cunha 333, 12500-000 Guaratinguet\'a, SP, Brazil.}
\maketitle
\begin{abstract}
{\small{In this work we report new results concerning the question of dynamical
mass generation in the Lorentz and PCT violating quantum elec\-tro\-dy\-na\-mics.
A one loop calculation  for the  vacuum polarization tensor is presented. The electron 
propagator, ``dressed'' by a Lorentz breaking extra term in the fermion Lagrangian density, 
is approximated by its first order: this scheme is shown to break gauge invariance. Then we 
rather consider a full calculation, to second order in the Lorentz breaking parameter: we 
recover gauge invariance and use the Schwinger-Dyson equation to discuss the full photon 
propagator. This allows a discussion on a possible photon mass shift as well as measurable, 
observable physical consequences, such as the Lamb-shift.
}}
%
%
\end{abstract}
%
\renewcommand{\thefootnote}{\fnsymbol{footnote}} 
\def\0{\begin{equation}}
\def\1{\end{equation}}
\def\2{\begin{eqnarray}}
\def\3{\end{eqnarray}}
\def\>{{\rangle}}
\def\<{{\langle}}
\def\H{{\rm {H}}}
\def\p{{\bf {p}}}
\def\la{{\lambda}}
\def\LA{{\Lambda}}
\def\al{{\alpha}}
\def\te{{\theta}}  
\def\be{{\beta}}	
\def\ka{{\kappa}}
\def\ga{{\gamma}}  
\def\Ga{{\gamma}}
\def\de{{\delta}}  
\def\De{{\Delta}}
\def\si{{\sigma}}  
\def\Si{{\Sigma}}
\def\om{{\omega}}  
\def\Om{{\Omega}}
\def\nf{{\infty}}  
\def\nl{{\newline}}
\def\ra{{\longrightarrow}}
\def\beq{{\begin{equation}}}
\def\eeq{{\end{equation}}}
\def\beqa{{\begin{eqnarray}}}  
\def\dst{{\displaystyle}}
\def\eeqa{{\end{eqnarray}}}
\def\nnb{{\nonumber}}


\newpage


\section{Introduction}

The PCT theorem states that if a field theory satisfies the following axioms: 
{\it a) locality, b) Lorentz invariance and c) analiticity of the Lorentz group
representation  in the boost parameters}, the PCT transformation is a symmetry of
the theory itself \cite{WI64}. In this context, the invariance under the Lorentz 
group represents one  of the fundamental axioms in the construction of a relativistic 
quantum field theory,  which includes the minimal $SU(3) \times SU(2) \times U(1)$ 
Standard Model. 

The possibility of nature to reveal a small violation of the Lorentz and PCT 
symmetries has been object of an intense research which includes different areas
of physics, ranging from Quantum Optics to Neutrino Physics \cite{KO99}. However,
up to now, there is no conclusive experimental evidence for the violation  of
axiom {\it b)}.

From the theoretical standpoint, there arose a controversy on a possible 
Chern-Simons-like term generated through radiative corrections in an extended
version  of Quantum Electrodynamics (QED) \cite{CO99} - \cite{SY03}. In fact, 
some authors  claimed that the addition to the QED Lagrangian density of a
renormalizable and Lorentz violating term such as \footnote{where :
\begin{itemize}
\item $b^{\alpha}$ is a set of four constants, which selects a preferencial 
direction in space-time, thus violating Lorentz invariance;

\item $\psi$ represents the electron field;  

\item $\gamma_5$ a hermitian matrix with the properties $\{\gamma_5,\gamma_\alpha\}=0$ 
and ${\rm tr}\gamma_5 \gamma^{\alpha} \gamma^{\beta} \gamma^{\gamma} \gamma^{\delta} = 
4 i \epsilon^{\alpha \beta \gamma \delta}$.
\end{itemize}}
\0
\label{e1} {\cal L}_{LB} = -b^{\alpha}{\bar \psi} \gamma_{\alpha} \gamma_5 \psi
\1
\noindent induces a Chern-Simons-like term, namely, 
\0
\label{e2} 
{\cal L}_{CS} = \frac{1}{2} c_{\mu} \epsilon^{\mu \alpha \beta \gamma}
F_{\alpha \beta} A_{\gamma}.
\1
\noindent thus violating the Lorentz and PCT symmetry. {\it Note that such a term 
(\ref{e2}) does not destroy the gauge invariance of the action.} 

\noindent As a matter of facts, at least two questions arose :
\begin{itemize}
\item As soon as Lorentz symmetry is violated, one could wonder about the validity
of any calculation : should one consider - as done for example in \cite{AGS} - that 
the loop momenta phase space is no longer Lorentz covariant? Should the regularization 
preserve Lorentz invariance? Here, we shall follow the conservative point of view 
initiated by Colladay and Kostelecky in which the sole breaking of Lorentz invariance 
comes from the added Lagrangian (\ref{e1}).

\item The Lagrangian (\ref{e1}) modifies the fermion propagator and two approches are 
possible. Do an expansion in $b^{\alpha}$ of the propagator which amounts to consider 
(\ref{e1}) as a new interaction: we call this approch a perturbative one. Either try 
to compute with the complete $b^{\alpha}$ dependent fermion propagator: we call this 
approach a non-perturbative one.
\end{itemize}

Then, if the Lorentz breaking is the minimal one and the electron propagator
expanded in powers of $b^{\alpha}\,,$ and if the theory is correctly defined 
through Ward identities and normalisation conditions, no Chern-Simons term 
appears \cite{BO01}\footnote{This is due to the fact that such a term (\ref{e2}) 
is a kind of minor modification of the gauge fixing term, and then no renormalised. 
Then, if absent at the classical level, it remains absent at the loop level.}. 
A similar result  was first obtained in the analysis due to Coleman and Glashow
\cite{CO99} and this result was afterwards confirmed through a proper-time approach 
by Sytenko and Rulik \cite{SY03}.

In this work we intend to reexamine the role of gauge invariance and the issue of its 
possible breaking; moreover, we want to study a possible dynamical mass generation 
for the photon. So, we compute the vacuum polarization amplitude $\Pi^{\mu \nu}$ in QED 
(regulated by means of the gauge invariant Pauli-Villars-Rayski (P-V-R) scheme \cite{RA48}) 
when the tiny Lorentz breaking term  (\ref{e1}) is added.

First, in Section 2, and in the spirit of \cite{LU04}, we consider the loop expansion 
of the vacuum polarization tensor when the fermion propagator that results from the 
QED Lagrangian density modified by term (\ref{e1}) is the upper mentionned  "non 
perturbative one". However, to simplify the discussion and thanks to the smallness of 
parameter $b^{\alpha}$, we approximated the infinite series in $b^{\alpha}$ by its 
linear contribution, to each order in the $\hbar$ perturbative series. This procedure 
essentially consists in dressing the fermion propagator with linear corrections in 
$b^{\alpha}$. However, this scheme suffers from many difficulties and we do not pursue 
its study.

Second, in a more standard way, in Section 3 we consider (\ref{e1}) as an interaction 
term and then do a double expansion in $b_{\alpha}$ and $\hbar$. We give the complete 
one loop calculation to second order in the breaking parameter $b_{\alpha}\;.$ As expected 
from general results \cite{BO01} we check that gauge invariance is recovered and, if an 
adequate normalization condition is chosen, the photon may remain massless.

Then, in Section 4, the full photon propagator will be obtained by summing the perturbative 
series through the Schwinger-Dyson Equation \cite{DY49}, allowing  for a detailed 
discussion on the dynamical mass generation (the pole shifting in the photon propagator). 
Consequences on the Lamb-shift are also addressed.

Finally, some concluding remarks are offered in Section 5.

\section{The One Loop Vacuum Polarization Tensor in Extended QED}

In the extended  version of QED, defined by the Lagrangian density 
\0
\label{e3} {\cal L} = -\frac{1}{4}F_{\mu\nu}F^{\mu\nu} - 
\frac{1}{2\alpha}(\partial_{\mu}A^{\mu})^2 + {\bar \psi} 
(i {\slash \!\!\! \partial} - e {\slash \!\!\!\! A} -  
{\slash \!\!\! b}\gamma_5  - m ) \psi,
\1

\noindent the fermion propagator is
\2 
S(l) = \frac{i}{ {\slash \!\!\! l} - m - {\slash \!\!\! b} \gamma_5 } =
\sum_{n=0}^{\infty}\frac{i}{ {\slash
\!\!\! l} - m }\left\{  -i \ {\slash
\!\!\! b} \gamma_5\frac{i}{ {\slash \!\!\! l} - m }\right\}^n  =
\sum_{n=0}^{\infty} S_n (l) ,\nnb 
\3 
and its linear approximation, used in that section, writes:
\0
\label{e4} S^L (l) = S_0 (l) + S_1(l) = \frac{i}{ {\slash \!\!\! l} - m } -i
\frac{i}{ {\slash \!\!\! l} - m }
\ {\slash \!\!\! b} \gamma_5\frac{i}{ {\slash \!\!\! l} - m },
\1 
In that linear approximation, the one-loop vacuum polarization tensor is 
\0
\label{e5}
\Pi^{\mu \nu} (p, m, b) = -\; \!(-i \;e)^2 \int \frac{d^4 l}{(2 \pi)^{4}} \; {\rm
Tr}  [ \gamma^{\mu} S^L(l) \gamma^{\nu} S^L(l+p) ].
\1
$\Pi^{\mu \nu} (p, m, b)$ then admits the decomposition 
\0
\label{e6}
\Pi^{\mu \nu} (p, m, b) = \Pi^{\mu \nu}_{0} (p,m) + \Pi^{\mu \nu}_{b} (p,m,b) +
\Pi^{\mu \nu}_{bb} (p,m,b).
\1 
In the last expression,
\2
\Pi^{\mu \nu}_{0} (p,m) =  e^2 \int \frac{d^4 l}{(2 \pi)^{4}} \; {\rm Tr}  [
\gamma^{\mu} S_0 (l) \gamma^{\nu} S_0 (l+p) ] \nnb
\3 
is the usual QED vacuum polarization tensor, 
\2
\Pi^{\mu \nu}_{b} (p, m, b) =   e^2 \int \frac{d^4 l}{(2 \pi)^{4}} \; {\rm Tr} 
[ \gamma^{\mu} S_0 (l) \gamma^{\nu} S_1(l+p) + \gamma^{\mu} S_1(l) \gamma^{\nu}
S_0 (l+p) ]
\3 
is the linear $b_{\alpha}$ contribution, and 
\2
\Pi^{\mu \nu}_{bb} (p, m, b) =   e^2 \int \frac{d^4 l}{(2 \pi)^{4}} \; {\rm Tr} 
[ \gamma^{\mu} S_1(l) \gamma^{\nu} S_1(l+p) ]
\3 
is a ${\cal O}(b^2)$ contribution to the polarization tensor, whose thorough
calculation will be one of the focuses of the present work.

Diagrammaticaly, the expansion (\ref{e6}) may be represented by the series 
\begin{center} 
\begin{picture}(400,120)(0,0)
\Text(0,95)[]{\sl $\Pi^{\mu \nu} (p,m,b)  \equiv $}
\Photon(36,95)(63,95){2}{4}
\Text(43,110)[t]{\small \sl $p$}
\Line(44,90)(50,100)
\GOval(78,95)(7,15)(0){0.4}
\Photon(93,95)(121,95){2}{4}
\Text(108,110)[t]{\small \sl $p$}
\Line(107,90)(113,100)
\Vertex(63,95){1}
\Vertex(93,95){1}
\Text(63,85)[b]{\sl $\nu$}
\Text(93,85)[b]{\sl $\mu$}
\Text(133,95)[]{=} 
%
%
\Photon(142,95)(170,95){2}{4}
\ArrowArcn(180,95)(7,90,90)
\Text(180,115)[t]{\small \sl $l+p$}
\Line(149,90)(155,100)
\GOval(182,95)(7,15)(0){1}
\Photon(196,95)(225,95){2}{4}
\ArrowArcn(180,95)(7,280,280)
\Text(180,75)[b]{\small \sl $l$}
\Line(207,90)(213,100)
\Vertex(168,95){1}
\Vertex(196,95){1}
\Text(233,95)[]{+} 
%

\Text(20,40)[]{\sl $ + $}

\Photon(30,40)(60,40){2}{4}
\Line(43,35)(49,45)
\GOval(74,40)(7,15)(0){1}
\Text(74,60)[t]{\small \sl $l+p$}
\Text(74,30)[t]{\small \sl $l$}
\ArrowArcn(74,40)(7,90,90)
\Line(77,31)(73,35)
\Line(73,31)(77,35)
\Photon(88,40)(116,40){2}{4}
\Line(101,35)(107,45)
\Vertex(60,40){1}
\Vertex(88,40){1}

%
\Text(127,40)[]{+} 
%
%
\Photon(134,40)(164,40){2}{4}
\Line(147,35)(154,45)
\Text(179,60)[t]{\small \sl $l+p$}
\GOval(179,40)(7,15)(0){1}
\Text(179,30)[t]{\small \sl $l$}
\ArrowArcn(177,40)(7,280,280)
\Line(181,45)(177,49)
\Line(177,45)(181,49)
\Photon(194,40)(224,40){2}{4}
\Line(207,35)(213,45)
\Vertex(164,40){1}
\Vertex(194,40){1}

\Text(231,40)[]{+} 


\Photon(238,40)(268,40){2}{4}
\Line(251,35)(257,45)
\Text(284,60)[t]{\small \sl $l+p$}
\GOval(283,40)(7,15)(0){1}
\Text(284,30)[t]{\small \sl $l$}
\Line(285,45)(281,49)
\Line(281,45)(285,49)
\Line(285,31)(281,35)
\Line(281,31)(285,35)
\Photon(298,40)(328,40){2}{4}
\Line(311,35)(317,45)
\Vertex(268,40){1}
\Vertex(298,40){1}




\end{picture}
\end{center}
where
\begin{center} 
\begin{picture}(120,60)(0,0)

\Line(0,30)(30,30)
\Line(17,28)(13,32)
\Line(13,28)(17,32)
\Text(40,30)[]{\sl $\equiv$}
\Line(50,30)(80,30)
\Vertex(65,30){1}
\DashLine(65,30)(65,50){2}
\Line(67,48)(63,52)
\Line(63,48)(67,52)
\Text(79,57)[]{\small \sl $ b_{\alpha}$ }
\Text(65,20)[]{\small \sl $i \gamma^{\alpha} \gamma_5$ }
\Text(90,30)[]{\sl $+$}
\Text(110,30)[]{\sl $O( b^2 )$}
\end{picture}
\end{center} represents the linear $b_{\alpha}$ insertion to the (internal)
fermion lines.


By power counting, $\Pi^{\mu \nu}_{bb} (p, m, b)$, $\Pi^{\mu \nu}_{b} (p, m, b)$ 
and $\Pi^{\mu \nu}_{0} (p, m)$  are respectively logarithmicaly, linearly and
quadratically divergent in the ultra-violet region. Of course, a regularization
procedure is needed: in order to preserve as much as possible the symmetries of 
the classical theory (\ref{e3}) ({\it i.e.} gauge invariance), the Pauli-Villars-Rayski 
(P-V-R) regularization prescription will be employed \cite{RA48}. In this scheme, 
auxiliary fermion masses satisfying specific conditions are introduced. The original 
theory is recovered at the end of the calculations, by taking arbitrary large values 
for the auxiliary masses. 

Let us recall also that thanks to gauge invariance, the degree of divergence of
the tensor $\Pi^{\mu \nu}_{0}(p)$, identical to the pure QED sector, is reduced 
from 2 to 0 (logarithmic divergence).

To be consistent with the P-V-R scheme, the tensor $\Pi^{\mu \nu} (p,m,b)$ must be 
regulated as a whole object, which implies the replacement of expression
(\ref{e6}) by the sum
\0
\label{e66}
\Pi^{\mu \nu} (p,b) = \sum_{i=0}^{N} c_i \left( \Pi^{\mu \nu}_{0} (p, m_i) + 
\Pi^{\mu \nu}_{b} (p, m_i, b) + \Pi^{\mu \nu}_{bb} (p, m_i, b) \right),
\1 
where each term in the r.h.s. of (\ref{e66}) retains the original functional
form, except the fact that $m \rightarrow m_i$ (note also that we consider 
the same Lorentz breaking parameter $b^{\alpha}$ for all fermions). By analysing 
the structure of (\ref{e66}), it can be shown that, for auxiliary masses satisfying 
the conditions:
\2 
a) & \sum_{i=0}^N c_i & = 0  , \label{e67} \\ 
b) & \sum_{i=0}^N c_i m_i^2  & = 0  ,\nnb
\3 
the divergences in $\Pi^{\mu \nu} (p,m,b)$ all disappear. In the  above expressions, 
$c_0 = 1$ and $m_0 = m$ is the electron mass.

The results may be written as\footnote{Details on that calculation will be published elsewhere.} 
\2
\Pi^{\mu \nu}_{0} (p, m_i) &=& [g^{\mu \nu}p^2 -  p^{\mu} p^{\nu}] \Pi_{0} (p^2, m_i), \\
\Pi^{\mu \nu}_{b} (p, m_i, b) &=& \epsilon^{\mu \nu \alpha \beta} p_{\alpha} b_{\beta } \Pi_b (p^2, m_i) ,\\
\Pi^{\mu \nu}_{bb} (p,m_i,b) &=& {\mathbb{A}}((p.b)^2,b^2,p^2,m_i) g^{\mu\nu} 
+ {\mathbb B}(p^2,m_i) b^{\mu} b^{\nu} + \nnb \\ 
&+& {\mathbb C}(p^2,m_i) (p.b)(b^{\mu} p^{\nu} + b^{\nu} p^{\mu}) 
+ {\mathbb D}((p.b)^2,b^2,p^2,m_i) p^{\mu} p^{\nu}, 
\3
where
\2
\label{r1}
\Pi_{0} (p^2, m_i) & = &  i \frac{ e^2}{12 \pi^2} 
\left\{ \log\frac{m_i^2}{m^2}  -  p^2 \int_0^1 dz \;\frac{[1- 2 Z - 8 Z^2]}{2\Delta_i} \right\}, \nnb \\
\Pi_b (p^2, m_i) & = &  i\frac{e^2}{2 \pi^2} \left\{ p^2 \int_0^1 dz\; \frac{Z}{\Delta_i} \right\},\nnb\\ 
{\mathbb A}((p.b)^2,b^2,p^2, m_i) & = &-i \frac{e^2}{\pi^2} \left\{ \frac{b^2}{6}\log\frac{m_i^2}{m^2} + 
\int_0^1 dz \left[ \frac{-1 -4 Z + 8 Z^2}{12 \Delta_i} \; b^2 p^2 + \frac{ Z^2}{\Delta_i} (b.p)^2 +
\right. \right. \nnb \\   
& + & \left. \left. \frac{Z^2}{2[\Delta_i]^2} p^2[(b \cdot p)^2 -b^2 p^2] \right] \right\} , \\ 
{\mathbb B}(p^2, m_i) & = &-i\frac{ e^2}{\pi^2} \left\{ -\frac{1}{6}\log\frac{m_i^2}{m^2} + p^2 \; 
\int_0^1 dz \left[ \frac{1 + 10 Z - 8 Z^2}{12\Delta_i} +  \frac{Z^2}{[\Delta_i]^2} p^2 \right] \right\} , \nnb\\ 
{\mathbb C}(p^2, m_i) & = & -i\frac{e^2}{\pi^2}\left\{ \int_0^1 dz \left[  -\frac{Z(1-2 Z)}{\Delta_i} - 
\frac{Z^2}{[\Delta_i]^2} p^2 \right] \right\},\nnb\\ 
{\mathbb D}((p.b)^2,b^2,p^2, m_i) &=& -i\frac{ e^2}{\pi^2}  \left\{\int_0^1 dz \left[\frac{3 Z^2}{\Delta_i}b^2 +  
2\frac{Z^3}{[\Delta_i]^2}((p.b)^2+ b^2 p^2) \right] \right\} .\nnb
\3
In those expressions, we have set
\2 Z &=& z (1-z), \nonumber \\
\Delta_i & = & m_i^2 - z(1-z)p^2 \nonumber .
\3
Moreover, to simplify our results we used some integration by parts on $z$ and the
symmetry of the integral when $z$ is changed into $(1-z)$ : this allows us to rewrite 
the integrands as functions of $Z.$

A few comments are in order :

\begin{itemize}
\item The scale in the logarithms has been (arbitrarily) chosen to be the
electron mass, thanks to the condition (\ref{e67}-a) $\dst\sum_{i=0}^{i=N} c_i =
0\; .$

\item Since $\Pi^{\mu \nu}_{b}$ vanishes at $p^2 = 0$, no Chern-Simons-like term 
appear \cite{BO01}.

\item The vacuum polarization tensor is no longer transverse as :
\2
p_{\nu}\Pi^{\mu\nu} = [{\mathbb A} +(p.b)^2 {\mathbb C} + p^2 {\mathbb D}]p^{\mu} + 
(p.b)[{\mathbb B} + p^2 {\mathbb C}]b^{\mu} \nonumber 
\3 
and neither of the two square brackets vanish (just look at the logarithms). Moreover, 
no local counterterm can be added in order to recover transversality. Then gauge
invariance is definitely broken in that scheme. This might be a consequence of 
our specific expansion in $b_{\alpha}$ and, indeed, in the scheme of section 3, 
we shall recover gauge invariance at the one loop level by introducing new missing
Feynman graphs, which ammounts to consider all possible permutations among
photon and $b_{\alpha}$ linear insertions as, for example, in the case of the
gauge-invariant (finite) result for the ${\alpha}^2$ amplitude of Delbrück
scattering in ordinary QED. 
 
\item The "infinite" renormalization of the vacuum polarization tensor then
requires the following counterterm : 
\2
\label{ct1}
-\frac{e^2}{12 \pi^2} \left[ \sum_{i=0}^{i=N} c_i \log \frac{m_i^2}{m^2} \right] 
\left\{ \frac{1}{4}F_{\mu\nu}F^{\mu\nu} - b^2 A_{\mu}A^{\mu} + (b_{\mu}A^{\mu})^2 \right\}
\3 
The ${\cal O}(b^2)$ part of this counterterm is absent in the classical Lagrangian 
density, breaks gauge invariance and moreover seems to generate a $b$ dependent mass 
for the photon.

\item Once the previously defined "infinite" renormalization has been done, note that, 
contrarily to $\Pi^{\mu \nu}_{0}$ and $\Pi^{\mu \nu}_{b} $, $\Pi^{\mu \nu}_{bb} (p,m_i,b)$ 
does not vanish at $p^2 = 0$ . Denoting as $\Pi^{(R)\mu \nu}$ the renormalized vacuum 
polarization tensor\footnote{ Once the counterterm (\ref{ct1}) has been added, the limit 
$m_i \ra \infty$ may be safely taken, except for $i=0$ where $m_0 = m$ and $c_0 =1$}, one gets :

$$ \Pi^{(R)\mu \nu} _{p^2=0} = $$
$$ = - i \frac{e^2}{\pi^2}\int_0^1 dz \left[ \frac{ Z^2}{m^2} (b.p)^2  g^{\mu \nu} - 
\frac{Z(1-2 Z)}{m^2}(b.p) ( b^{\mu} p^{\nu} + b^{\nu} p^{\mu})  + (\frac{3 Z^2}{m^2} b^2 +
\frac{2 Z^3}{m^4}(b.p)^2) p^{\mu} p^{\nu} \right] = $$
$$= -i \frac{e^2}{10 m^2 \pi^2}\left[\frac{1}{3}(b.p)^2  g^{\mu \nu} - (b.p) 
(b^{\mu} p^{\nu} + b^{\nu} p^{\mu}) +(b^2 +\frac{(p.b)^2}{7 m^2}) p^{\mu} p^{\nu}
\right] $$

\noindent Then, if one tried to enforce the normalization condition 
$$ \Pi^{(R)\mu \nu} _{p^2=0} = 0\;,$$ a necessary condition for the photon to
remain massless, an extra finite counterterm would be required but, due to the
last term, it would have a canonical dimension 6,
\0
\frac{e^2}{70 (m^2)^2 \pi^2} b^{\alpha} b^{\be}
(\partial_{\alpha}\partial_{\mu}A^{\mu})(\partial_{\be}\partial_{\nu}A^{\nu}) \nonumber
\1  
which breaks renormalizability.
\end{itemize}

All this rules out this approximation scheme and we now consider the usual "perturbative approach".

\section{The Vacuum Polarization Tensor in Extended QED: Gauge Invariant Approach}

If we no longer restrict ourselves to the linear approximation for the fermion
propagator, two new Feynman graphs have to be added in the calculation of
$\Pi^{\mu \nu}_{bb} (p, m, b)\; :$
\2
\tilde{\Pi}^{\mu \nu}_{bb} (p, m, b) = e^2\,\int \frac{d^4 l}{(2 \pi)^{4}}
\left\{ {\rm Tr}  [ \gamma^{\mu} S_2(l) \gamma^{\nu} S_0(l+p) ]+{\rm Tr}  
[\gamma^{\mu} S_0(l) \gamma^{\nu} S_2(l+p) ]\right\}
\3

Diagrammaticaly, the new graphs are represented by (the series) 

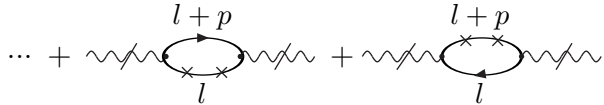
\begin{center} 
\begin{picture}(140,70)(0,0)

\Text(05,40)[]{\sl $ ... $}

\Text(20,40)[]{\sl $ + $}

\Photon(30,40)(60,40){2}{4}
\Line(43,35)(49,45)
\GOval(74,40)(7,15)(0){1}
\Text(74,60)[t]{\small \sl $l+p$}
\Text(74,30)[t]{\small \sl $l$}
\ArrowArcn(74,40)(7,90,90)
\Line(70,32)(66,36)
\Line(66,32)(70,36)
\Line(83,32)(79,36)
\Line(79,32)(83,36)
\Photon(88,40)(116,40){2}{4}
\Line(101,35)(107,45)
\Vertex(60,40){1}
\Vertex(88,40){1}

%
\Text(127,40)[]{+} 
%
%
\Photon(134,40)(164,40){2}{4}
\Line(147,35)(154,45)
\Text(179,60)[t]{\small \sl $l+p$}
\GOval(179,40)(7,15)(0){1}
\Text(179,30)[t]{\small \sl $l$}
\ArrowArcn(177,40)(7,280,280)
\Line(175,44)(171,48)
\Line(171,44)(175,48)
\Line(187,44)(183,48)
\Line(183,44)(187,48)
\Photon(194,40)(224,40){2}{4}
\Line(207,35)(213,45)
\Vertex(164,40){1}
\Vertex(194,40){1}

\end{picture}
\end{center}

The result may be written as \footnote{Details on that calculation will be published elsewhere.} 
\2
\tilde{\Pi}^{\mu \nu}_{bb} (p,m_i,b)& = &\tilde{{\mathbb{A}}}((p.b)^2,b^2,p^2,m_i)
g^{\mu\nu} + \tilde{{\mathbb B}}(p^2,m_i) b^{\mu} b^{\nu}  + \nnb\\ &+ &
\tilde{{\mathbb C}}(p^2,m_i) (p.b)( b^{\mu} p^{\nu} + b^{\nu} p^{\mu}) +
\tilde{{\mathbb D}}((p.b)^2,b^2,p^2,m_i) p^{\mu} p^{\nu}, 
\3 
where
\2
\tilde{{\mathbb A}}((p.b)^2,b^2,p^2, m_i) & = & -i \frac{e^2}{\pi^2} \left\{
-\frac{b^2}{6}\log \frac{m_i^2}{m^2} + \int_0^1 dz \left[ 
\frac{1 -8 Z - 8 Z^2}{12\Delta_i} b^2 p^2 + \frac{4 Z^2}{\Delta_i} (b.p)^2 +
\right. \right. \nnb \\   & + & \left. \left. \frac{2 Z^3}{[\Delta_i]^2} p^2(b
\cdot p)^2 - \frac{Z^2}{2[\Delta_i]^2} (p^2)^2 b^2  \right] \right\} , \\
\tilde{{\mathbb B}}(p^2, m_i) & = & -i\frac{ e^2}{\pi^2} \left\{
\frac{1}{6}\log \frac{m_i^2}{m^2} -\int_0^1 dz
\left[ \frac{(1 -4 Z)(1+2 Z)}{12\Delta_i} p^2 \right] \right\} , \\
\tilde{{\mathbb C}}(p^2, m_i) & = & -i\frac{e^2}{\pi^2}\left\{ \int_0^1 dz \left[ 
-\frac{2 Z^2}{\Delta_i} \right] \right\},\\
\tilde{{\mathbb D}}((p.b)^2,b^2,p^2, m_i) &=& -i\frac{ e^2}{\pi^2}  \left\{
\int_0^1 dz \left[ \frac{Z(1-Z)}{2\Delta_i}b^2 +  \frac{Z^2(3-4 Z)}{4[\Delta_i]^2} b^2 p^2 -
2\frac{Z^3}{[\Delta_i]^2}(p.b)^2 \right] \right\} .
\3 
As in Section 2, our results were simplified thanks to integration by parts on $z$ 
and the symmetry of the integral when $z$ is changed into $(1-z)$ and we have set
\2 
Z &=& z (1-z), \nonumber \\
\Delta_i & = & m_i^2 - z(1-z)p^2 \nonumber .
\3

When these results are combined with those in Section 2, the full tensor 
$\bar{\Pi}^{\mu \nu}_{bb} (p, m, b) \;$ simplifies to :
\2
\label{r3}
\bar{\Pi}^{\mu \nu}_{bb} (p, m, b) = -i \frac{e^2}{\pi ^2} \left\{ - X^{\mu\nu} +
Y^{\mu\nu}\right\} \int_0^1 dz \left[ \frac{Z}{\Delta_i}  + \frac{ Z^2}{[\Delta_i]^2} p^2 \right]  ,
\3 
where we have introduced the two transverse tensors of dimension 4, quadratic
in $b^{\alpha}\;:$
\2
\label{r4}  
a) & X^{\mu\nu}  = & b^2 (g^{\mu\nu}p^2 - p^{\mu}p^{\nu})  \nnb \\ 
b) & Y^{\mu\nu}  = & g^{\mu\nu}(p.b)^2 + p^2 b^{\mu}b^{\nu} -(p.b)(p^{\mu}b^{\nu}
+p^{\nu}b^{\mu})  .
\3 
Again, a few comments are in order:
\begin{itemize}
\item With that double expansion in $b^{\alpha}$ and $\hbar\,,$ gauge invariance
holds at the regularized level and then we also checked that {\it no ultra-violet
divergence remains} in the full $\bar{\Pi}^{\mu \nu}_{bb}$. Then the solely
required "infinite" renormalization of the vacuum polarization tensor is the
standard QED one 
\2
\label{cta} 
-\frac{e^2}{12 \pi^2} \left[ \sum_{i=0}^{i=N} c_i \log \frac{m_i^2}{m^2} \right] 
\left\{ \frac{1}{4}F_{\mu\nu}F^{\mu\nu} \right\}.
\3

\item Since $\Pi^{\mu \nu}_{b}$ vanishes at $p^2 = 0$, no Chern-Simons-like term 
appear \cite{BO01}.

\item However, $\bar{\Pi}^{\mu \nu}_{bb} (p,m_i,b)$ still does not vanish at $p^2 = 0 .$ 
Denoting as $\bar{\Pi}^{(R)\mu\nu}$ the renormalized vacuum polarization tensor (again, 
once the QED counterterm has been added, the limit $m_i \ra \infty$ may be taken, except
for $i=0$ where $m_0 = m$ and $c_0 =1$), one gets:
 
$$ \bar{\Pi}^{(R)\mu \nu} _{p^2=0} = -i \frac{e^2}{6\pi^2 m^2} \left [g^{\mu\nu}(p.b)^2 + 
b^2 p^{\mu}p^{\nu} -(p.b)(p^{\mu}b^{\nu} + p^{\nu}b^{\mu}) \right]. $$

Then, we have two possibilities: 
\begin{itemize}
\item in a kind of "minimal subtraction" (let us recall that the infinite subtraction has 
been done (\ref{cta})), this non-vanishing behaviour might indicate a non vanishing mass 
for the photon - despite the validity of gauge invariance. In subsection 4.2, we shall 
analyse this possibility through the Schwinger-Dyson equation;

\item in  a more standard way, one may enforce the normalization condition 
$$ \bar{\Pi}^{(R)\mu \nu} _{p^2=0} = 0 $$ in order to the photon remain massless 
( see subsection 4.1): then an extra finite counterterm is required. Among those 
possible which compensate the previous quantity up to $p^2 = 0$ vanishing contributions 

$$ i \frac{e^2}{6\pi^2 m^2} \left[ g^{\mu\nu}(p.b)^2 + b^2 p^{\mu}p^{\nu} - (p.b)(p^{\mu}b^{\nu} + 
p^{\nu}b^{\mu}) \right] +{\cal O}(p^2) = $$

$$= i \frac{e^2}{6\pi^2 m^2}\left[Y^{\mu\nu} -  X^{\mu\nu}\right] +{\cal O'}(p^2),$$

we choose the unique gauge invariant term:
\2
\label{ctb} \frac{e^2}{6 \pi^2 m^2}\left[g^{\mu\nu}(b^{\alpha} F_{\alpha\mu})(b^{\be}
F_{\be\nu})  -\frac{b^2}{4}F_{\mu\nu}F^{\mu\nu} \right] .
\3

\end{itemize}

\end{itemize}

\noindent Let us now discuss more precisely the consequences of the Lorentz
breaking on the photon mass and the Lamb-Shift.

\section{Physical consequences of the Lorentz breaking}

In order to obtain the full photon propagator, the Schwinger-Dyson equation must 
be solved \cite{DY49}. In a given order of the perturbative series, it implies the
summation of an infinite set of proper Feynman diagrams. In this way, as
\2
({G_0}^{-1})^{\mu \nu} = i [ p^2 g^{\mu \nu} - (1 - \frac{1}{\alpha}) p^{\mu} p^{\nu} ] \nnb
\3
is the inverse of the QED free photon propagator, the  Schwinger-Dyson equation
\footnote{Recall that the renormalized tensor $\bar{\Pi}^{(R)\mu \nu} (p, m, b)$
is obtained after addition of the infinite and eventually of the finite counterterms 
(resp (\ref{cta}) and (\ref{ctb})): then the limit $m_i \ra \infty,\;i \neq 0\;, $ 
may be taken.}
\2 
({G}^{-1})^{\mu \nu} = (G_0^{-1})^{\mu \nu} - \bar{\Pi}^{(R)\mu \nu} (p, m, b) \nnb
\3 
leads us to 
\2
\label{551} ({G}^{-1})^{\mu \nu} = i\left[A' g^{\mu \nu} + B' b^{\mu} b^{\nu} + C' 
(b^{\mu} p^{\nu} + b^{\nu} p^{\mu})  + D' p^{\mu} p^{\nu} + E' \epsilon^{\mu \nu
\alpha \beta } p_{\alpha} b_{\beta }\right] ,
\3 
where the values of the functions $A', B', C', D'$ and $E'$ depend on the finite 
subtraction (the normalisation condition).

From (\ref{551}) we obtain the inverse tensor:
\2
\label{55333} G_{\mu \nu} =-i\left[ A'' g_{\mu \nu} + B'' b_{\mu} b_{\nu} + C'' ( b_{\mu}
p_{\nu} + b_{\nu} p_{\mu})  + D'' p_{\mu} p_{\nu} + E'' \epsilon_{\mu \nu \alpha
\beta } p^{\alpha} b^{\beta }\right],
\3 
where\footnote{We have defined:
$$ Deno1 = A'^2 + [p^2 b^2 -(p.b)^2]E'^2 ,$$
$$ Deno2 = A'[A' + b^2 B' +2(p.b) C' + p^2 D'] + [p^2 b^2 -(p.b)^2](B'D' - C'^2).$$ 
Note that if $({G}^{-1})^{\mu \nu} $ were purely transverse ($A'+(p.b) C' + p^2 D' = 
(p.b) B' +p^2 C' = 0$), the inversion would not be possible as $Deno2 \equiv 0 .$}
\2
A'' & = & \frac{A'}{Deno1}, \nnb \\ 
B'' & = & \frac{p^2 E'^2}{A'[Deno1]} - \frac{A' B' + p^2(B' D' - C'^2)}{A'[Deno2]}, \nnb \\ 
C'' & = & - \frac{(p.b) E'^2}{A'[Deno1]} - \frac{A' C' - (p.b)(B' D' - C'^2)}{A'[Deno2]}, \label{554} \\ 
D'' & = & \frac{b^2 E'^2}{A'[Deno1]} - \frac{A' D' + b^2(B' D' - C'^2)}{A'[Deno2]}, \nnb \\ 
E'' & = & - \frac{E'}{Deno1} ,\nnb
\3
We now discuss successively  the case when a normalisation condition is imposed as well 
as the ``minimal subtraction'' scheme.

\subsection{ With the Normalization Condition $ \bar{\Pi}^{(R)\mu \nu} _{p^2=0} = 0$ }
In this case:
\2
A'_{NC} & = & p^2\left\{1+ \frac{e^2}{2 \pi^2}\left[p^2 \beta(p^2,m^2) + 
[(p.b)^2 - p^2 b^2]\chi(p^2,m^2)\right]\right\} , \nnb\\ 
B'_{NC} & = & \frac{e^2}{2 \pi^2}(p^2)^2 \chi(p^2,m^2) ,\nnb\\ 
C'_{NC} & = & -\frac{e^2}{2 \pi^2}(p^2)(p.b) \chi(p^2,m^2), \label{5522}\\ 
D'_{NC} & = & \frac{1-\alpha}{\alpha} + \frac{e^2}{2 \pi^2}[ -p^2 \beta(p^2,m^2) + 
p^2 b^2\chi(p^2,m^2)] ,\nnb \\ 
E'_{NC} & = & -\frac{e^2}{2 \pi^2} p^2 \gamma(p^2,m^2) \nnb,
\3 
where:
\2
\beta(p^2,m^2) & = &  \int_0^1 dz \;\frac{[1- 2 Z - 8 Z^2]}{12\Delta_0} \nnb\\
\chi(p^2,m^2) & = & \frac{2}{p^2} \int_0^1 dz \left\{ \left[\frac{Z}{\Delta_0}  +
\frac{ Z^2}{[\Delta_0]^2} p^2 \right] - \left[ \frac{Z}{m^2}  \right]\right\} = 
2 \int_0^1 dz\; Z^2 \left[ \frac{1}{m^2 \Delta_0} + \frac{1}{[\Delta_0]^2} \right] 
\label{5533}\\ 
\gamma(p^2,m^2) & = & \int_0^1 dz \; \frac{Z}{\Delta_0}. \nnb
\3 
The functions $\beta,\;\chi$ and $\gamma $ are finite at $p^2=0$:

$$\beta(0,m^2)=1/(30 m^2)\;,\; \chi(0,m^2)=2/(15 m^4)\;,\;\gamma(0,m^2)=1/(6 m^2)\;.$$ 

We obtain:
$$ Deno1 \simeq {A'}_{NC}^2 \simeq (p^2)^2 \left\{ 1 + \frac{e^2}{\pi^2} \left[ p^2 \beta(p^2,m^2) + 
[(p.b)^2 - p^2 b^2] \chi(p^2,m^2) \right] \right\}$$

$$ Deno2 \simeq \frac{(p^2)^2}{\alpha} \left[ 1 + \frac{e^2}{2\pi^2} p^2 \beta(p^2,m^2) \right]\;.$$

As a consequence: 
\2 
A'' & \simeq & \frac{1}{p^2}\left\{1- \frac{e^2}{2 \pi^2}\left[p^2 \beta(p^2,m^2)
+[(p.b)^2 -p^2 b^2]\chi(p^2,m^2)\right]\right\} \nnb\\  
B'' & \simeq & -\frac{e^2}{2\pi^2} \chi(p^2,m^2) ,\nnb\\  
C'' & \simeq & \frac{e^2}{2\pi^2}\frac{(p.b)}{(p^2)} \chi(p^2,m^2) ,\label{562}\\  
D'' & \simeq & -\frac{1}{(p^2)^2}\left\{1-\alpha+ \frac{e^2}{2 \pi^2}\left[-p^2 \beta(p^2,m^2) + p^2
b^2\chi(p^2,m^2)\right]\right\} ,\nnb \\ 
E'' & \simeq & \frac{e^2}{2 \pi^2}\frac{1}{p^2} \gamma(p^2,m^2))\;. \nnb
\3

So, we have checked that, as a consequence of the normalization condition 
$ \bar{\Pi}^{(R)\mu \nu} _{p^2=0} = 0,$ the photon remains massless. 

If one expands the vacuum polarisation tensor around $p^2 = 0$, one obtains a 
new contribution to the Lamb-shift. We consider a static configuration where 
$p^2 = - \vec{p}^2$ and extract from (\ref{562}) an approximate expression for 
the Coulomb interaction:
$$\frac{e^2}{\vec{p}^2}\left[1+ \frac{e^2 \vec{p}^2}{60 \pi^2 m^2}\left\{1 - 
4\left[\frac{(\vec{p}.\vec{b})^2}{m^2\vec{p}^2} + \frac{ b^2}{m^2}\right]\right\}\right]\;.$$
As a consequece, the effect of Lorentz breaking on that part of the Lamb-shift 
is drastically tiny (recall that $ |b/m| < 10^{-22}$ (\cite{BL04})).

\subsection{``Minimal Subtraction'' Case}

If only the infinite counterterm (\ref{cta}) is added, one obtains:
\2
A'_{MS} & = & p^2 + \frac{e^2}{2 \pi^2}\left[(p^2)^2 \beta(p^2,m^2) + 
[(p.b)^2-p^2 b^2]\phi(p^2,m^2)\right] , \nnb\\ 
B'_{MS} & = & \frac{e^2}{2 \pi^2} p^2 \phi(p^2,m^2) ,\nnb\\ 
C'_{MS} & = & -\frac{e^2}{2 \pi^2} (p.b) \phi(p^2,m^2),\label{552}\\ 
D'_{MS} & = & \frac{1-\alpha}{\alpha} + \frac{e^2}{2 \pi^2}[-p^2 \beta(p^2,m^2) + 
b^2\phi(p^2,m^2)] ,\nnb \\ 
E'_{MS} & = & -\frac{e^2}{2 \pi^2} p^2 \gamma(p^2,m^2) \nnb,
\3 
where:
\2
\beta(p^2,m^2) & = &  \int_0^1 dz \;\frac{[1- 2 Z - 8 Z^2]}{12\Delta_0} \nnb\\
\phi(p^2,m^2) & = & 2 \int_0^1 dz \left[\frac{Z}{\Delta_0}  + \frac{
Z^2}{[\Delta_0]^2} p^2 \right]  \label{553}\\
\gamma(p^2,m^2) & = & \int_0^1 dz\; \frac{Z}{\Delta_0} . \nnb
\3 
The functions $\beta(p^2,m^2),\phi(p^2,m^2)$ and $\gamma(p^2,m^2)$ are finite at $p^2 = 0\;:$
$$\beta(0,m^2)=1/(30 m^2)\;,\; \phi(0,m^2)=1/(3 m^2)\;,\;\gamma(0,m^2)=1/(6 m^2)\;.$$
We now obtain:
$$ Deno1 \simeq {A'}_{MS}^2 \simeq p^2 \left\{ p^2 + \frac{e^2}{\pi^2} \left[ (p^2)^2 \beta(p^2,m^2) +
[(p.b)^2 - p^2 b^2] \phi(p^2,m^2) \right] \right\}$$
$$ Deno2 \simeq \frac{(p^2)^2}{\alpha} \left\{ 1 + \frac{e^2}{2 \pi^2} p^2 \beta(p^2,m^2) \right\}.$$
As a consequence, a  shift of the zero mass pole seems to happen. Indeed, if one expands  
$Deno1$ and $Deno2$ near $p^2 = 0$, one gets  
$$ Deno1 \simeq p^2 \left\{ p^2 + \frac{e^2}{\pi^2} \left[ (p.b)^2 \phi(0,m^2)] + 
{\cal O}(p^2)\right] \right\} \simeq p^2 [p^2 + \mu^2] \;\;,\;\; Deno2 \simeq \frac{(p^2)^2}{\alpha}$$ 
with $$ \mu^2 \simeq  \frac{e^2}{\pi^2} \frac{(p.b)^2}{3 m^2}\;.$$
However, this ``tachyon-like'' singularity is not so simple to interpret due to its $p$ 
dependence and we shall not pursue that direction.

\section{Concluding Remarks}
In this paper we presented two calculations for the 1-loop vacuum polarization tensor in 
the Lorentz and PCT violating QED defined by (\ref{e1}). This was done using the canonical 
perturbation  theory and the P-V-R regularization scheme. In a first approximation we consider 
the electron propagator as ``dressed' up to first order in $b$. This scheme breaks gauge 
invariance and we argue it should not be considered. In a second calculation, we consider 
the complete expansion to second order in $b$. Gauge invariance is then recovered and we 
discussed the  normalisation conditions that might be added in order the photon to remain 
massless. Let us remark that our calculation for the $\Pi_{bb}^{\mu\nu}$ part allows us to 
correct a sign error in a previous calculation (formula (20) of \cite{Chen}). 

The full photon propagator is then obtained through the Schwinger-Dyson equation allowing 
a detailed  analysis of its singular structure. In particular, we found that in a ``minimal 
scheme'' without normalisation condition, a new tachyonic-like pole appears. On the contrary, 
if the vacuum polarisation tensor is normalised such that it vanishes at $p^2 = 0$, no spurious 
pole appears and the photon remains massless. The effect on the Lamb-shift comming from the 
dependence on $b$ of the vacuum polarisation appears to be quite negligible.

Further investigations concerning the full 1-loop renormalization of the extend QED as well as 
its thermal version is now in progress and will be reported in a forthcoming paper \cite{TO05}.
%
%
\vspace{1.0cm} 

\noindent {\bf Acknowledgements.} 
LCC and JLT are grateful to the LPTHE - Universit\'e de Paris VI et VII, where part of 
this work was developed. LCC acknowledges, in special, Prof. Olivier Babelon, Director of 
the LPTHE, and Profs. B. Diu, B. Dou\c cot and B. Delamotte, for the kind hospitality. 
LCC also thanks Fapesp-Brazil and CNRS-France for financial support. JLT thanks FAPESP-Brazil 
for the financial support.

\bibliographystyle{plain}

\end{document}